 \documentclass[final,3p,times]{elsarticle}
\usepackage{amssymb}

\journal{Powder Technology}

\begin{document}

\begin{frontmatter}

\title{Multiscale modeling of rapid granular flow with a hybrid discrete-continuum method}

\author[label1,label2]{Xizhong Chen}
\author[label1]{Junwu Wang\corref{cor1}}
\cortext[cor1]{Corresponding author}
\ead{jwwang@ipe.ac.cn}
\author[label1]{Jinghai Li}
\address[label1]{State Key Laboratory of Multiphase Complex Systems, Institute of Process Engineering, Chinese Academy of Sciences, Beijing 100190, P. R. China}
\address[label2]{University of Chinese Academy of Sciences, Beijing, 100049, P. R. China}

\begin{abstract}
Both discrete and continuum models have been widely used to study rapid granular flow, discrete model is accurate but computationally expensive, whereas continuum model is computationally efficient but its accuracy is doubtful in many situations. Here we propose a hybrid discrete-continuum method to profit from the merits but discard the drawbacks of both discrete and continuum models. Continuum model is used in the regions where it is valid and discrete model is used in the regions where continuum description fails, they are coupled via dynamical exchange of parameters in the overlap regions. Simulation of granular channel flow demonstrates that the proposed hybrid discrete-continuum method is nearly as accurate as discrete model, with much less computational cost.
\end{abstract}


\end{frontmatter}


\section{Introduction}
\label{}
Granular matter which consists of macroscopic particles can widely be found in nature and in industry \citep{jaeger1996granular,yang1998fluidization}. A better understanding of granular matter is not only desirable for physicists, but also for engineers from various sectors, such as mining, pharmaceutical and chemical industries \citep{andreotti2013granular}. Extensive theoretical, numerical and experimental studies have been devoted to this fascinating area \citep{goldhirsch2003rapid,aranson2006patterns,luding2009towards}, however, our understanding is still far from satisfactory after several decades' efforts due to its dissipative, non-linear and non-equilibrium characteristics, reflected by the lack of a general theory for describing its hydrodynamics \citep{goldhirsch2003rapid,kadanoff1999built}.

To simulate granular flow, both macroscopic continuum model and microscopic discrete model have been extensively used \citep{brilliantov2004kinetic,poschel2005computational}. Continuum model, solving the conservation equations of mass, momentum and energy, is very useful for analyzing and designing industrial processes involving a large number of discrete particles. However, it is much less mature compared with the classical fluid mechanics theory for molecular gas or liquid fluid, mainly due to the dissipative nature of particle-particle interactions and the resultant lack of scale separation \citep{goldhirsch1999scales, goldhirsch2003rapid}, and the formation of heterogeneous structures, such as particle clustering structure \citep{goldhirsch1993clustering}. Continuum method may also be inadequate in situations when no accurate boundary condition can be formulated or due to the existence of Knudsen layer \citep{campbell1993boundary,galvin2007role}. On the other hand, discrete model \citep{cundall1979discrete} tracks the motion of each particle according to Newton's law, and accordingly, provides detailed information about the dynamics of granular flow. Unfortunately, discrete model is computationally extremely demanding for engineering applications.

In this study, a hybrid discrete-continuum method is developed for modelling granular flow, taking the merits of both continuum and discrete methods but discarding their drawbacks. The idea is to concurrently couple physical descriptions at different scales to solve the dilemma that has been described, as in the hybrid atomistic-continuum methods for rarefied gas flows \citep{wadsworthone1990hybrid}, dense fluids \citep{kalweit2008multiscale,mohamed2010review} and solid mechanics \citep{miller2009unified}. Continuum model is used to model the majority of simulation domain while discrete method is used within the domains where continuum description is inadequate.
\section{Method}\label{sec:method}
\subsection{Continuum method}
\begin{figure}
\centering
\includegraphics[scale=1.5]{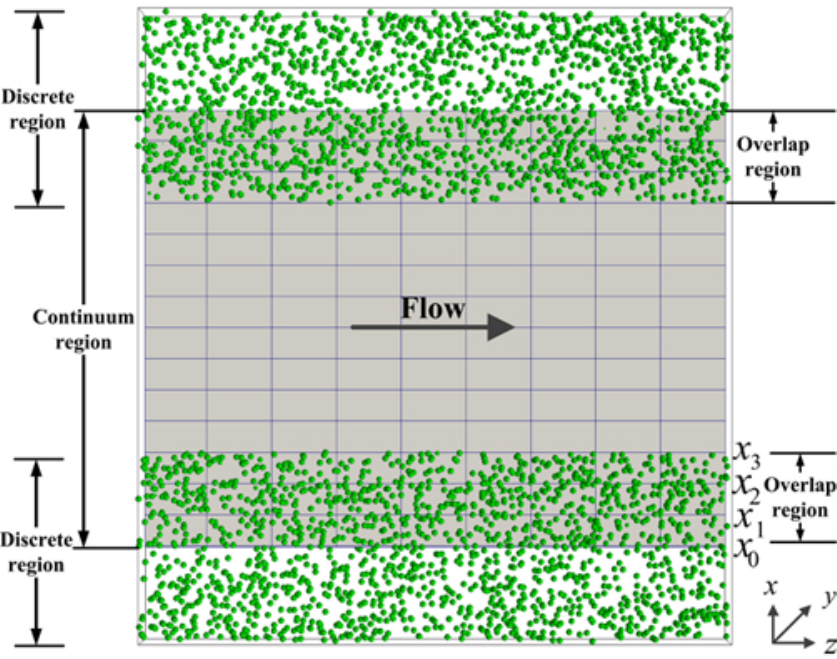}
\caption{\label{fig:1}  Schematic of the granular channel flow simulated by the hybrid discrete-continuum method.  The simulation domain is divided into a continuum region located at the center of the channel and two discrete regions near the two side walls. They coincidentally have two overlap regions and each overlap region is further divided into three parts: a discrete boundary layer $\left(x_{2}\sim x_{3}\right)$, a buffer layer $\left(x_{1}\sim x_{2}\right)$ and a continuum boundary layer $\left(x_{0}\sim x_{1}\right)$. }
\end{figure}

The hybrid discrete-continuum method is based on domain decomposition as illustrated in Fig.\ref{fig:1}.
The method has been used in the study of micro- and nano-fluid flow \citep{mohamed2010review,kalweit2008multiscale,wijesinghe2004discussion}. The simulation domain is divided into a continuum region located at the center of the channel and two discrete regions near the two side walls. Two overlap regions are constructed to ensure the continuity of mass and momentum in the entire simulation domain. In the continuum region, Navier-Stokes equation combined with kinetic theory of granular flow is applied \citep{lun1984kinitic}. The governing equations are numerically solved with finite difference method on a staggered grid \citep{kuipers1993computer,van2006multiscale}. The mass and momentum conservation equations are given as follows,
\begin{equation}
\frac{\partial }{{\partial t}}\left( {{\phi _s}{\rho _s}} \right) + \nabla  \cdot \left( {{\phi _s}{\rho _s}{{\bf{U}}_c}} \right) = 0
\end{equation}
\begin{equation}
\frac{\partial }{{\partial t}}\left( {{\phi _s}{\rho _s}{{\bf{U}}_c}} \right) + \nabla  \cdot \left( {{\phi _s}{\rho _s}{{\bf{U}}_c}{{\bf{U}}_c}} \right) =  - \nabla {p_s} + \nabla  \cdot {{\bf{\tau }}_s}
\end{equation}
where ${p_s}$ and ${{\bf{\tau }}_s}$  are the granular pressure and solid stress tensor, respectively. The stress tensor is linearly related to the rate-of-strain tensor,
\begin{equation}
{{\bf{\tau }}_s} = {\mu _s}\left( {\nabla {{\bf{U}}_c} + \nabla {{\bf{U}}_c}^T} \right) + \left( {{\lambda _s} - \frac{2}{3}{\mu _s}} \right)(\nabla \cdot{{\bf{U}}_c}){\bf{I}}
\end{equation}
where ${\mu _s}$ is shear viscosity and ${\lambda _s}$ is bulk viscosity. The particulate phase stresses are closed using kinetic theory of granular flow which solves a separate conservation equation for granular temperature.
\begin{equation}
\frac{3}{2}\left[ {\frac{{\partial \left( {{\phi _s}{\rho _s}{\Theta _s}} \right)}}{{\partial t}} + \nabla \cdot\left( {{\phi _s}{\rho _s}{{\bf{U}}_c}{\Theta _s}} \right)} \right] = \left( { - {p_s}{\bf{I}} + {{\bf{\tau }}_s}} \right):\nabla {{\bf{U}}_c} + \nabla \cdot\left({{k_s}\nabla {\Theta _s}}\right) - \gamma
\end{equation}
where ${k_s}$  is heat conductivity of granular phase given by:
\begin{equation}
{k_{_s}} = \frac{{150{\rho _s}{d_p}\sqrt {{\Theta _s}\pi } }}{{384\left( {1 + e} \right){g_0}}}{\left[ {1 + \frac{6}{5}{\phi_s}{g_0}\left( {1 + e} \right)} \right]^2} + 2{\rho _s}\phi_s^2{d_p}\left( {1 + e} \right){g_0}\sqrt {\frac{{{\Theta _s}}}{\pi }}
\end{equation}
The solids pressure, shear and bulk viscosity are expressed as:
\begin{equation}
{p_s} = {\phi_s}{\rho _s}{\Theta _s} + 2{\rho _s}\left( {1 + e} \right)\phi_s^2{g_0}{\Theta _s}
\end{equation}
\begin{equation}
{\mu _s} = \frac{4}{5}{\rho _s}{d_p}\phi_s^2{g_0}\left( {1 + e} \right)\sqrt {\frac{{{\Theta _s}}}{\pi }}  + \frac{{{\phi_s}{\rho _s}{d_p}\sqrt {\pi {\Theta _s}} }}{{6(3 - e)}}\left( {1 + \frac{2}{5}\left( {1 + e} \right)\left( {3e - 1} \right){\phi_s}{g_0}} \right)
\end{equation}
\begin{equation}
{\lambda _s} = \frac{4}{3}{\phi_s}{\rho _s}{d_p}{g_0}\left( {1 + e} \right)\sqrt {\frac{{{\Theta _s}}}{\pi }}
\end{equation}
The energy dissipation due to inelastic collision is expressed as:
\begin{equation}
\gamma  = \frac{{12\left( {1 - {e^2}} \right){g_0}}}{{{d_p}\sqrt \pi  }}{\rho _s}\phi_s^2\Theta _s^{{3 \mathord{\left/
 {\vphantom {3 2}} \right.
 \kern-\nulldelimiterspace} 2}}
\end{equation}
Within these formulations, ${g_0}$ is the radial distribution function that is given as follows:
\begin{equation}
{g_0} = {\left[ {1 - {{\left( {\frac{{{\phi_s}}}{{{\phi_{s,\max }}}}} \right)}^{{1 \mathord{\left/
 {\vphantom {1 3}} \right.
 \kern-\nulldelimiterspace} 3}}}} \right]^{ - 1}}
\end{equation}
The tangential velocity and granular temperature at the wall are calculated using Johnson and Jackson model \citep{johnson1987frictional}.
\begin{equation}
{u_{s,w}} =  - \frac{{6{\mu _s}{\phi _{s,\max }}}}{{\sqrt 3 \pi \varphi {\rho _s}{\phi _s}{g_0}\sqrt {{\Theta _s}} }}\frac{{\partial {u_{s,w}}}}{{\partial n}}
\end{equation}
\begin{equation}
{\Theta _{s,w}} =  - \frac{{{k_s}{\Theta _s}}}{{{\gamma _{s,w}}}}\frac{{\partial {\Theta _{s,w}}}}{{\partial n}} + \frac{{\sqrt 3 \pi \varphi {\rho _s}{\phi _s}u_{s,slip}^2{g_0}{\Theta _s ^{3/2}}}}{{6{\phi _{s,\max }}{\gamma _{s,w}}}}
\end{equation}
\subsection{Discrete method}
In the discrete regions, the linear spring-dashpot discrete element method \citep{cundall1979discrete,ye2004numerical} is used:
\begin{equation}
{m_a}\frac{{d{{\bf{u}}_p}}}{{dt}} = \sum\limits_{b \in contactlist} {{{\bf{F}}_{ab,n}}}
\end{equation}
In this model, the normal component of the contact force between two particles is calculated as follows:
\begin{equation}
{{\bf{F}}_{ab,n}} =  - {k_n}{\delta _n}{{\bf{n}}_{ab}} - {\eta _n}{{\bf{u}}_{ab,n}}
\end{equation}
where ${k_n}$ is the normal spring stiffness, ${\bf{n}}_{ab}$ is the normal unit vector,  ${\eta _n}$ is the normal damping coefficient and ${\bf{u}}_{ab,n}$ is the normal relative velocity. The overlap ${\delta _n}$ is given by:
\begin{equation}
{\delta _n} = ({R_a} + {R_b}) - \left| {{{\bf{r}}_b} - {{\bf{r}}_a}} \right|
\end{equation}
where ${R_a}$ and ${R_b}$ denote the radii of the interacting particles, ${{\bf{r}}_a}$ and ${{\bf{r}}_b}$ denote the position vector of the particles. The normal unit vector is defined as:
\begin{equation}
{{\bf{n}}_{ab}} = \frac{{{{\bf{r}}_b} - {{\bf{r}}_a}}}{{\left| {{{\bf{r}}_b} - {{\bf{r}}_a}} \right|}}
\end{equation}
The normal component of the relative velocity between particle a and particle b is
\begin{equation}
{{\bf{u}}_{ab,n}} = \left( {{{\bf{u}}_{ab}} \cdot {{\bf{n}}_{ab}}} \right){{\bf{n}}_{ab}}
\end{equation}
The normal damping coefficient is calculated as follows:
\begin{equation}
{\eta _n} = \frac{{ - 2\sqrt{{m_{ab}}{k_n}} \ln e}}{{\sqrt {{\pi ^2} + {{\ln }^2}e} }}
\end{equation}

Note that (i) we start with the simplest kinetic theory, wall boundary condition and contact model available in literature to develop the hybrid multi-scale model, while more advanced kinetic theory \citep{garzo1999dense}, wall boundary condition \citep{li2012revisiting,soleimani2015comparison} and contact model \citep{di2004comparison} can be implemented later. (ii) In addition to its simplicity of linear spring-dashpot discrete element method, it leads to a constant restitution coefficient, which is in agreement with the basic assumption of kinetic theory of granular flow used here   \citep{van2006multiscale, schwager2007coefficient}.

\subsection{Hybrid method}
The simulation domain in hybrid method is divided into a continuum region located at the center of the channel and two discrete regions near the two side walls. They coincidentally have two overlap regions. The design of overlap region is the most critical part of the hybrid method and should be as simple as possible in order to guarantee the high efficiency of hybrid method over pure discrete method. In present study, the overlap region is further divided into three parts: a discrete boundary layer, a buffer layer and a continuum boundary layer.
The boundary condition for continuum model is provided by discrete model through appropriately averaging the particle quantities as shown in the ${x_0} \sim {x_1}$ cell , i.e.
\begin{equation}
{\phi _J}{\rho _s} = \frac{1}{{{V_J}}}\sum\limits_{i \in J} {{m_i}}
\end{equation}
\begin{equation}
{\phi _J}{\rho _s}{{\bf{u}}_{J}} = \frac{1}{{{V_J}}}\mathop \sum \limits_{i \in J} {m_i}{{\bf{v}}_i}
\end{equation}
\begin{equation}
{\phi _J}{\rho _s}{\Theta _J} = \frac{1}{{3{V_J}}}\mathop \sum \limits_{i \in J} {m_i}\left( {{{\bf{v}}_i} - {{\bf{u}}_{J}}}  \right) ^2
\end{equation}
Where $i$ denotes a particular particle within the cell $J$, $m_i$ is its mass and ${{\bf{v}}_i}$ is its velocity, ${V_J}$ is the volume of cell $J$, ${\phi _J}$ is the solid volume fraction, ${\rho _s} $ is the particle mass density, ${{\bf{u}} _J}$ is the cell averaged particle velocity and  ${\Theta _J}$ is the granular temperature of cell $J$. Because discrete region is terminated in the $x_{2}\sim x_{3}$ cell, the missed information should be remedied by continuum method. Handling the boundary condition from continuum to discrete method is a more subtle issue than the reverse. The related issue is that the degrees of freedom in the discrete description of particles are much greater than those of the continuum one. Here the dynamics of particles in the $x_{2}\sim x_{3}$ cell are constrained to match the continuum model via constraint dynamics \citep{o1995molecular}. The new velocity of particle $i$ located within cell $J$ is therefore given by:
\begin{equation}
{{\bf{\dot x}}_i} = {{\bf{v}}_i} + \xi \left( {{{\bf{u}}_c} - \frac{1}{{{N_J}}}\mathop \sum \limits_{i = 1}^{{N_J}} {{\bf{v}}_i}} \right)
\end{equation}
Where $\bf{u}_c$ is the velocity of particle calculated by continuum method in the specific cell. The constraint strength $\xi$ is introduced to relax the particle momenta to the local continuum value and here we use a value of unity \citep{cosden2013hybrid,nie2004continuum}.

To maintain the mass conservation in discrete domain, a ghost wall and a flux monitor are placed at the termination of discrete domain $\left( x_{3}\right)$. Once a particle overlaps with the ghost wall, particle-wall collision will prevent the particle from freely drifting away from the discrete domain. The actual mass flux across the ghost wall is achieved through the flux monitor based on the local continuum flow fields. The number of particles to be inserted into or deleted from cell $J$ in a continuum interval is calculated by:
\begin{equation}
n = ({{\bf{A}}_J} \cdot {{\bf{u}}_c}){\rm{ }}{\phi _J}{\rho _s}\Delta {t_c}/{m_i}
\end{equation}
Where ${{\bf{A}}_J} \cdot {{\bf{u}}_c}$ is the normal flux of the cell $J$ calculated by the continuum method and $\Delta {t_c}$ is the time step of continuum method. $n$ is always rounded to integer since only a whole particle can be added or removed. The residue of particle is recorded and accumulated for the next event. If $n$ is negative, it indicates that mass flow out of discrete domain into continuum domain and thus $n$ particles closest to $x_{3}$ are deleted. Otherwise, if $n$ is positive then $n$ particles should be added. The new inserted particles are randomly located in the interface nearest $x_{3}$ provided that they do not overlap with the existed particles. Note that currently the way to insert particles may not work well in very dense granular flow and other more sophisticated methods should be implemented \citep{van2010particle,markesteijn2011connecting}. The velocities of newly inserted particle are generated from Maxwell distribution with mean and standard deviation determined by the local continuum velocity and granular temperature as:
\begin{equation}
f({\bf{x}},{{\bf{v}}_i}) = \frac{1}{{{{(2\pi {\Theta _c})}^{3/2}}}}\exp \left[ { - \frac{{{{({{\bf{v}}_i} - {{\bf{u}}_c})}^2}}}{{2{\Theta _c}}}} \right]
\end{equation}
Where ${{\bf{u}}_c}$ and  $\Theta_c$ are the average velocity and granular temperature calculated by continuum model. To minimize the effect of data exchange, a buffer layer lies between the continuum boundary layer and the discrete boundary layer is constructed. Note that a simpler hybrid method for homogeneous granular flow has been presented in \cite{chen2015hybrid}.

\section{Results}\label{sec:results}
We have selected the granular channel flow as an example to demonstrate the concept of hybrid discrete-continuum method. The reason is that the validation range of continuum model is not a priori known for general granular flow, but for channel flow studied here we could know in advance the places where continuum model fails, that is, continuum model validates in the center region but fails close to the walls. This choice significantly simplifies the problem we studied, because we don't need a criterion to determine the places of overlap region. On the other hand, it has been shown that wall boundary condition has a great impact on the global behaviour of granular flow, it influences the stress state in its immediate vicinity and then propagates to the entire flow field \citep{campbell1993boundary}. This phenomenon occurs on the order of few particle sizes or within the Knudsen layer which usually cannot be handled by continuum model \citep{campbell1993boundary,galvin2007role}. Therefore, it is also an excellent example to highlight the merits of hybrid discrete-continuum method.

In the simulation of granular channel flow, due to the lack of scale separation of rapid granular flow \citep{goldhirsch2003rapid}, the theoretical foundation underlying the average of DEM results for the boundary condition of continuum method is much less solid than that of ordinary gases, the calculated averaged properties may depend on the the number of particles used, therefore, in this study, two dimensional equations are solved in continuum region, while discrete simulations are three-dimensional with periodic boundary condition in y-direction, in order to increase the sample of particles for a better averaging. Particles are assumed to be smooth, uniform, and spherical. The diameter of particles denoted as $d$ is $1.2mm$ and the density is $2000kg/{m^3}$. The depth, length and height of the channel are $6.1d$, $100d$ and $666d$, respectively. The domain is divided into $20 \times 64$ structural cells. In hybrid simulation, the discrete regions span $0 \sim 35d$ (7 numerical cells) and $65 \sim 100d$, the continuum region spans $15 \sim 85d$ in $x$ direction. Thus, each overlap region has $30d$ (or 6 cells) in $x$ direction. Quantities sampled from the first 2 cells nearest to the pure discrete domain are used to provide boundary condition for continuum method due to the used staggered computational cell which means the scalar and vectorial variables are not saved at the same location but have half-cell dislocation, the next 3 cells are used as buffer layer and the last cell is the discrete boundary in which the particle dynamics are constrained by the local continuum state as have been described above. A zero gradient condition is assigned to the top outlet and the Johnson and Jackson boundary condition \citep{johnson1987frictional} is applied at the wall in the continuum simulation, where the particle-wall coefficient of restitution $e_w$ is the same as the particle-particle coefficient of restitution $e$ and the specularity coefficient $\varphi$ is zero.

In discrete simulations, the particle-particle and particle-wall coefficient of normal restitution are the same as in continuum model and the coefficient of tangential restitution is unity to achieve the simulation of smooth particle \citep{poschel2005computational}. The stiffness of the linear spring model is $3000 N/m$ and simulation results are no longer sensitive to its further increase. The simulation time step is $1.0\times10^{-5}s$ and the statistics are collected over two million time steps after an initial million time steps.To highlight the merits of hybrid discrete-continuum method, each case is simulated using continuum method (CM), discrete method (DM) and hybrid method (HM), and the results of DM are used as the benchmark data for comparison.

\subsection{channel flow with uniform inlet velocity}
\begin{figure} %figure* and scale 2.0 for full page
\centering
\includegraphics[scale=1.5]{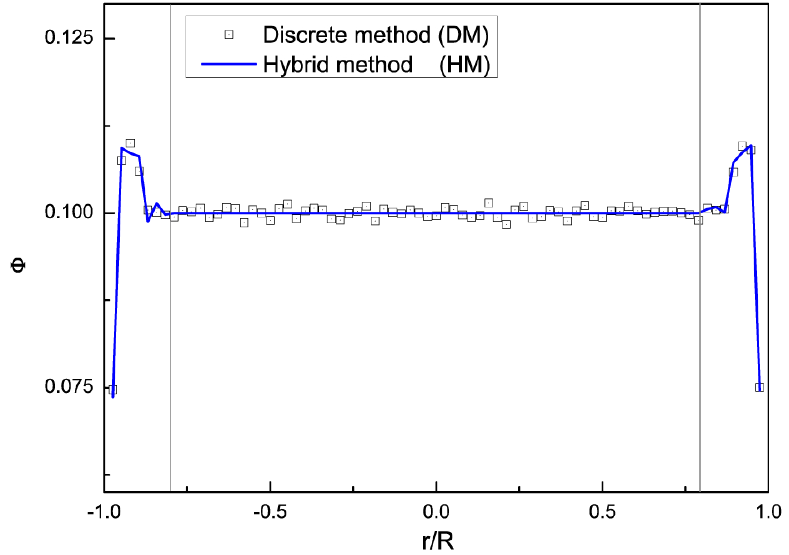}
\caption{\label{fig:3}
  Radial distribution of mean solid volume fraction at the dimensionless height of 0.6, the part of the channel between the vertical lines is simulated using continuum method. The coefficient of restitution $e=e_w=0.95$.}
\end{figure}

In order to validate the hybrid discrete-continuum method, we begin with the simplest homogeneous case. Particles are inserted into the channel continuously from the bottom with a uniform inlet velocity $\left(2.0m/s\right)$ and the inlet solid volume fraction $\phi$ is 0.1.
\begin{figure} %figure* and scale 2.0 for full page
\centering
\includegraphics[scale=1.5]{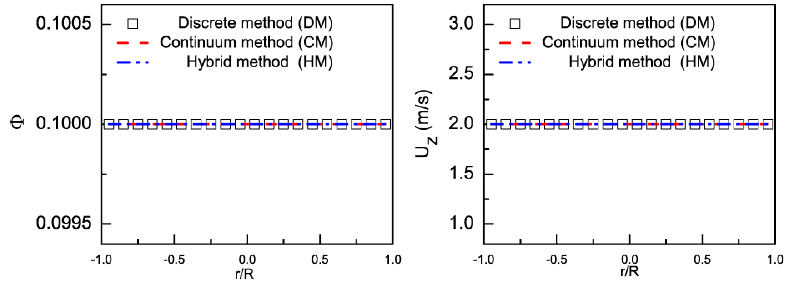}
\caption{\label{fig:2}
  Radial distribution of mean solid volume fraction and particle velocity at the dimensionless height of 0.6.}
\end{figure}
Fig.\ref{fig:2} shows the results of radial solid volume fraction and particle velocity cross the channel. A perfect agreement between the results of all three methods is observed (the relative differences are below ${10^{ - 6}}$). However, although both CM and DM predict a homogeneous solid volume fraction cross the channel, it is well known that there is oscillation in solid concentration at about 5 particle diameters away from the channel wall \citep{benenati1962void}. To explore this phenomenon, we decrease the binning width used to calculate the mean value of solid concentration to $1.3d$ and reanalysis the simulation results of the discrete simulation. As expected, the solid volume fraction do oscillate near the wall as shown in Fig.\ref{fig:3}, which has also been reported by \cite{hartkamp2012study} and MRI measurement in a cylindrical container \citep{sederman2001structure}. Note that the details of oscillation depend on the size of bins used to calculate the mean value and smoother data can be obtained using a Gaussian kernel \citep{goldhirsch2010stress,hartkamp2012study}. While the oscillations exist across the whole channel in DM, the magnitude is significant smaller in the center of the channel than the near wall regions. The solid volume fraction in the center converges to a bulk density since the effect of walls in this region disappears. Thus, the part of the channel between the vertical lines as shown in Fig.\ref{fig:3} is simulated using continuum method. The near wall oscillations phenomenon is captured by HM in nature and this is expected to be much more important for gas-solid multiphase system and the heat transfer behaviour between solid particles and wall in fluidized beds \citep{tsuji2013effect}.

\subsection{channel flow with Parabolic inlet velocity}
\begin{figure} %figure* and scale 2.0 for full page
\centering
\includegraphics[scale=1.5]{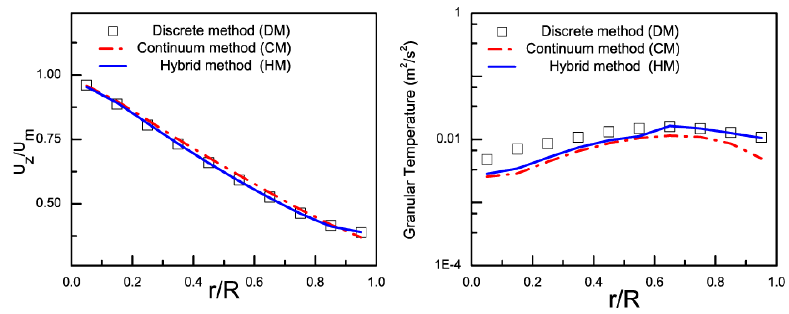}
\caption{\label{fig:5}
 Mean particle velocity and granular temperature at the dimensionless height of 0.6. The coefficient of restitution is $e=e_w=0.95$.}
\end{figure}

In this part, particles are inserted into the channel continuously from the bottom with a parabolic velocity ${U_z} = {U_m}\left( {1 - \frac{{{r^2}}}{{{R^2}}}} \right)$, where $U_m$ is the maximum particle velocity $\left(2.0m/s\right)$ in the center of the channel $\left(r=0\right)$, $R$ is half width of the channel and the inlet solid volume fraction $\phi$ is 0.1.
\begin{figure} %figure* and scale 2.0 for full page
\centering
\includegraphics[scale=1.5]{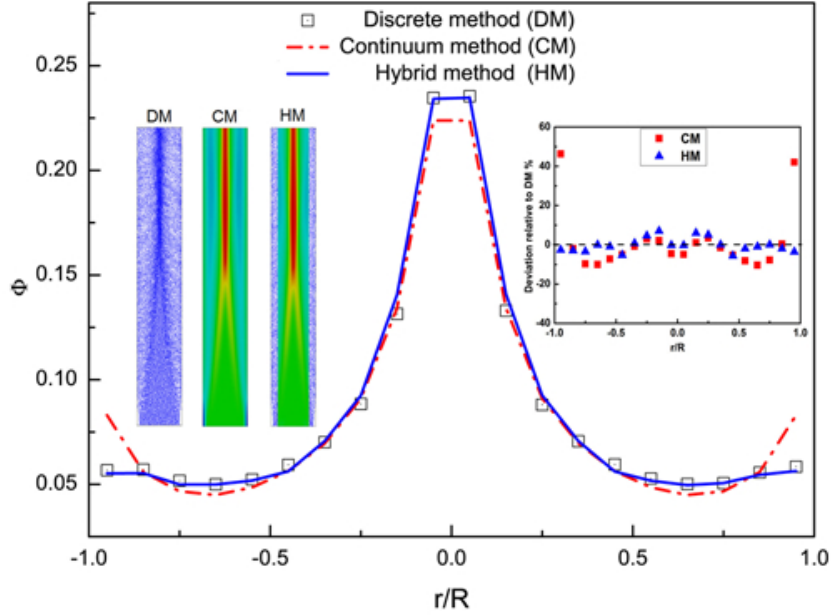}
\caption{\label{fig:4}
  Radial distribution of mean solid volume fraction at the dimensionless height of 0.6. The left inset shows the snapshots obtained using discrete method (DM), continuum method (CM) and hybrid method (HM), the right inset shows quantitative relative deviation.}
\end{figure}
Fig.\ref{fig:4} shows the radial distribution of mean solid volume fraction at the dimensionless height of 0.6 with $e=e_w=0.95$. The solid volume fraction distribution becomes inhomogeneous with the development of the flow, with a high density in the center and low density near the walls, as visualized in the left inset, the flow behaviour is qualitatively agreed with previous studies \citep{wang1997instabilities,liu2006parametric}. The quantitative deviation relative to DM, defined as $\frac{{\phi_{HM}} - {\phi_{DM}}}{{\phi_{DM}}}$ or $\frac{{\phi_{CM}} - {\phi_{DM}}}{{\phi_{DM}}}$, is shown in the right inset. It can be seen that the simulation results from HM and DM are in a good agreement (within $10\%$ deviations), however, large deviations can be found in the near wall region between CM and DM.  This is due to the fact that the range of the influence of particle-wall interaction is occurring over non-continuum scales (only spanning several particle diameters) as in previous studies \citep{campbell1993boundary,galvin2007role}. The thickness of the Knudsen-layer is approximate $8d$ where large deviations (about $46\%$) is observed between CM and DM.  However, although the particle-wall interaction only influences the stress state in its immediate vicinity, its effect can propagate to the entire flow field and finally result in a relatively large deviation at other parts of the simulation domain \citep{campbell1993boundary}, for example, the relative deviation at $\frac{r}{R}\approx0.7$ are larger than $10\%$. Note that the agreement between HM and DM may be improved if a fluctuating continuum model is used to replace current continuum model, as evident in the counterpart of hybrid method for molecular flow \citep{de2006multiscale,markesteijn2014concurrent}. Fig.\ref{fig:5} shows the mean particle velocity  and granular temperature across the channel at the dimensionless height of 0.6. Although both CM and DM predict a similar trend about the radial granular temperature profile, there is a relative large deviation between CM and DM as also observed in previous studies \citep{vescovi2014plane,almazan2013numerical}. The granular temperature in the center part of the channel predicted by HM does not show a significant improvement although the particle velocity and granular temperature simulated by HM are closer to those of DM as shown in Fig.\ref{fig:5}.

\begin{figure}
\centering
\includegraphics[scale=1.5]{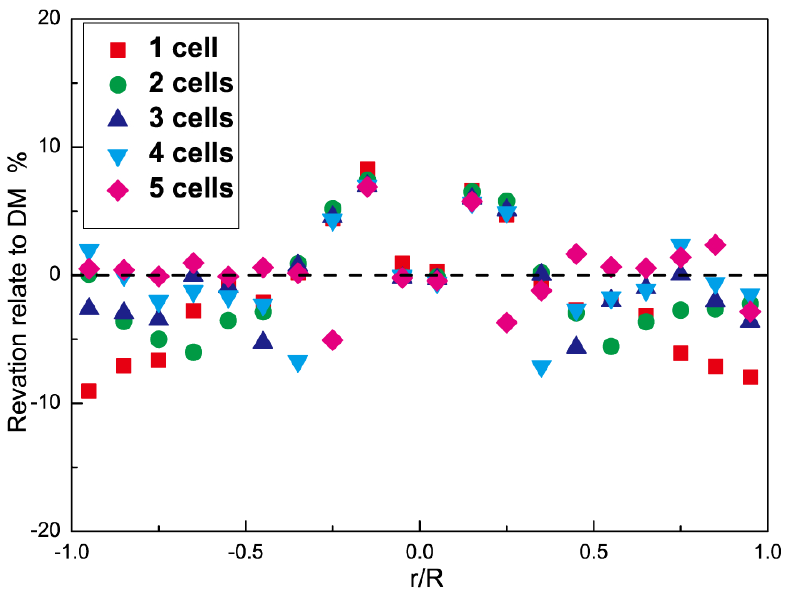}
\caption{\label{fig:6} The effect of cell numbers used for the buffer layer on the relative deviation of simulated solid volume fraction using hybrid method (HM).}
\end{figure}

Introduction of the buffer layer increases computational expense and thus should be as small as possible, while it should also be large enough to decay sufficiently the disturbance caused by the dynamical data exchange. The effect of cell numbers used for the buffer layer is shown in  Fig.\ref{fig:6}.
It seems that there is no clear rule, the extent of deviation not only depends on the number of cells used for buffer layer but also on the locations. All data are within $10\%$, indicating the rationality of the proposed method for data exchange. We have selected 3 cells for the buffer layer with certain degree of arbitrariness and also in view of the relative large deviation at near region in case of one used cell.

\begin{figure}
\centering
\includegraphics[width=12.7cm]{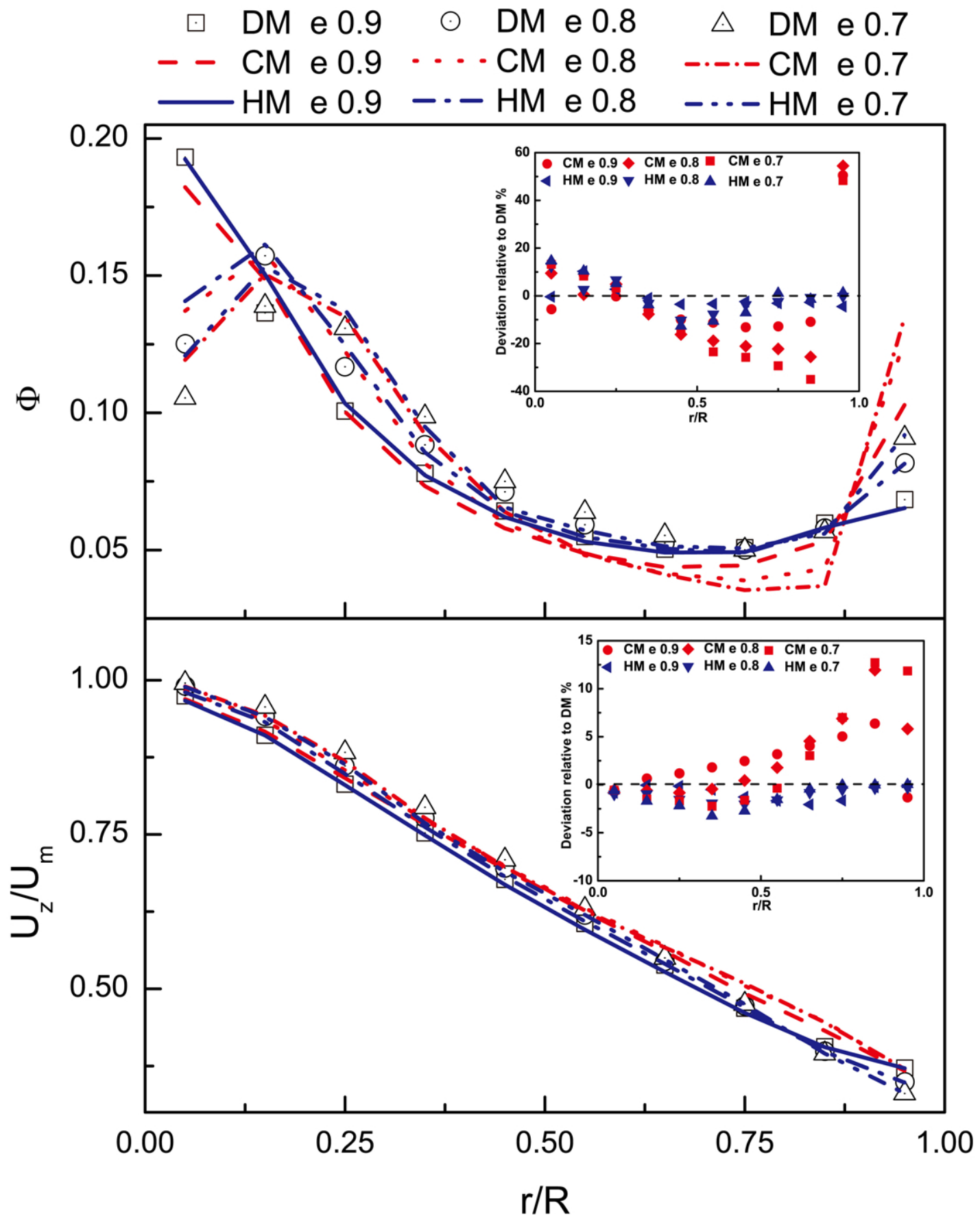}
\caption{\label{fig:7} Radial distribution of mean solid volume fraction and mean particle velocity at the dimensionless height of 0.6. The coefficients of restitution are $e=e_w=0.9$, $e=e_w=0.8$ and $e=e_w=0.7$.}
\end{figure}

The dissipative nature of particle-particle collision makes granular flow different from ordinary gases. Kinetic theories of granular flow used here are based on the qualitative assumption of small energy dissipation during particle-particle collision, that is, $1-e\approx0$, but it is unclear  what the  range of quantitative validation is. From the comparison between the results obtained from HM and DM as shown in Fig.\ref{fig:4}, we can infer that a value of $e=0.95$ gives an excellent agreement between CM and DM, if the particle-wall interactions can be correctly captured by CM. We therefore further check its validation range by decreasing the value of $e$. The results are shown in Fig.\ref{fig:7}.
Only half of the radial distribution of mean particle velocity and solid volume fraction and their relative deviation are plotted because of the symmetrical geometry. It can be seen that the difference between CM and DM is increasing with decreasing of $e$. In case of $e=0.7$, the relative deviation of solid volume fraction and mean particle velocity are higher than $50\%$ and $10\%$, respectively. However, in general, the slip velocity at wall and solid volume fraction distribution near the wall are correctly predicted by utilizing HM. Therefore, the results of HM are in much better agreement with those of DM, the relative deviation of solid volume fraction and mean particle velocity are always less than $20\%$ and $5\%$, respectively. This conclusion indicated that continuum treatment of particles provides a good approximation of its discrete nature even at low value of $e$, that is, CM may beyond its nominal range of validity as has been concluded in previous studies \citep{wang2013comparison, mitrano2014kinetic, louge2014surprising}.
\begin{figure} %figure* and scale 2.0 for full page
\centering
\includegraphics[scale=1.5]{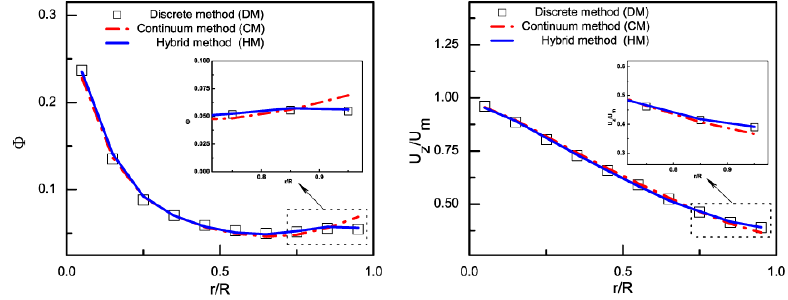}
\caption{\label{fig:8}
  Radial distribution of mean solid volume fraction and particle velocity, $e=0.95$,$e_w=1.0$, inlet solid volume fraction 0.1, $U_m=2.0m/s$.}
\end{figure}

\begin{figure} %figure* and scale 2.0 for full page
\centering
\includegraphics[scale=1.5]{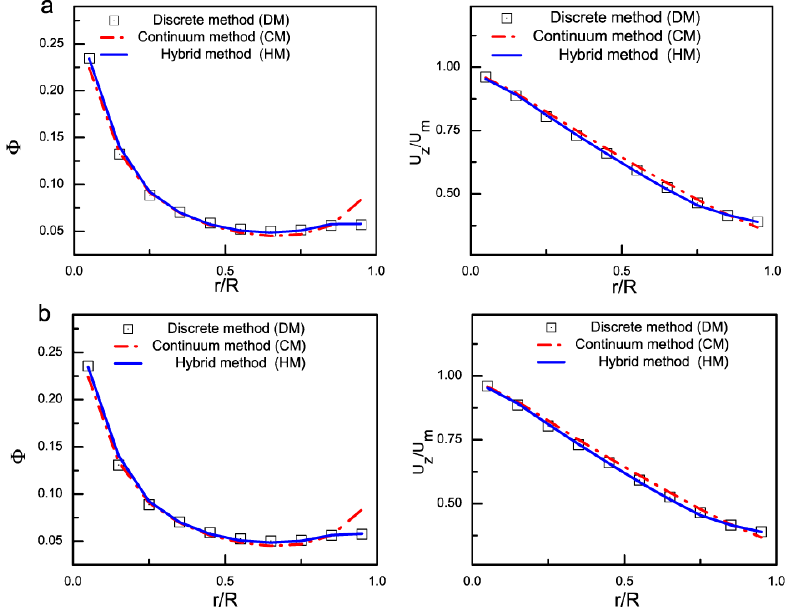}
\caption{\label{fig:9}
  Radial distribution of mean solid volume fraction and particle velocity, $e=e_w=0.95$,  inlet solid volume fraction 0.1, (a) $U_m=4.0m/s$, (b) $U_m=8.0m/s$.}
\end{figure}

\begin{figure} %figure* and scale 2.0 for full page
\centering
\includegraphics[scale=1.5]{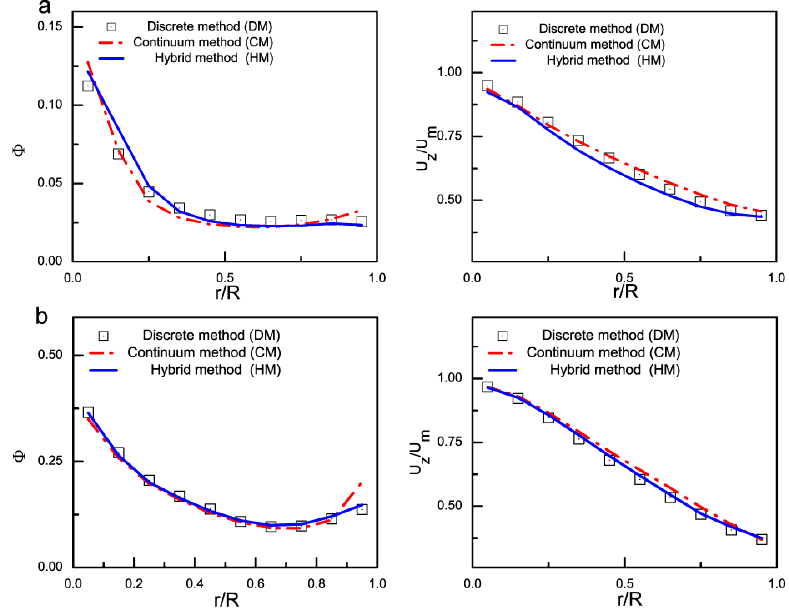}
\caption{\label{fig:10}
  Radial distribution of mean solid volume fraction and particle velocity, $e=e_w=0.95$, $U_m=2.0m/s$, inlet solid volume fraction (a) 0.05, (b) 0.2.}
\end{figure}

Fig.\ref{fig:8} shows an extreme case of elastic collisions of particles with the walls. It can be seen that there are still differences in the near wall region between the CM and DM, which indicates that the HM is necessary even in such an ideal situation. Furthermore, different flow parameters are simulated in order to explore the validity of the hybrid method. Fig.\ref{fig:9} shows radial distribution of mean solid volume fraction and particle velocity in the cases that the maximum inlet particle velocities are $4.0m/s$ and $8.0m/s$. Fig.\ref{fig:10} shows the results obtained from  the cases where inlet solid volume fractions equate 0.05 and 0.2. In all cases, a good agreement between HM and DM can be obtained.
Another attractive characteristic of the hybrid method is its computational efficiency compared with the discrete method for large scale granular flow. Even for the small scale problem studied here, the computation time is reduced to around $40\%$ of the full discrete method in a serial execution, and it is believed to result in a much more significant reduction of computational time when studying large scale problems that are the original motivation of present study. This sheds light on the potential of hybrid method for high accurate simulations of engineering problems.

\section{Conclusion}
We have proposed a hybrid discrete-continuum method for studying rapid granular flow. Simulations of granular channel flow demonstrated that the hybrid method can be nearly as accurate as discrete method due to its ability to capture the non-continuum particle-wall interactions, but its computational cost is much lower than discrete method. For a general granular flow, the places where the overlap region should be located are unclear, developing a criterion to identify the valid places of continuum description is critical for its success. Furthermore, the continuum model used here is only valid when there is no locally heterogeneous structures in the studied system, however, locally heterogeneous structures do exist in many granular systems \citep{aranson2006patterns}, for instance, particle clustering structure \citep{goldhirsch1993clustering}. Therefore, two more challenging problems that we faced are how to establish a continuum model that can account for the effects of the existence of locally heterogeneous structures, where the study of \cite{khain2007hydrodynamics} might be a nice source of idea, and how to determine the corresponding rules for the dynamical coupling of discrete and continuum models at overlap regions. Those will be the subjects of our future works.

\section{Acknowledgement}
We thank Professor Hans Kuipers at Eindhoven University of Technology for allowing the usage of his 3D-DPM code and Dr Lin Zhang at IPE, CAS for valuable discussions. Financial supports from National Natural Science Foundation of China (21422608 and 91334106) and Chinese Academy of Sciences (XDA07080200) are appreciated.

\section*{References}
\bibliographystyle{elsarticle-num-names}
\bibliography{hybrid}

\begin{thebibliography}{50}
\providecommand{\natexlab}[1]{#1}
\providecommand{\url}[1]{\texttt{#1}}
\providecommand{\urlprefix}{URL }
\expandafter\ifx\csname urlstyle\endcsname\relax
  \providecommand{\doi}[1]{doi:\discretionary{}{}{}#1}\else
  \providecommand{\doi}[1]{doi:\discretionary{}{}{}\begingroup
  \urlstyle{rm}\url{#1}\endgroup}\fi
\providecommand{\bibinfo}[2]{#2}

\bibitem[{Jaeger et~al.(1996)Jaeger, Nagel, and Behringer}]{jaeger1996granular}
\bibinfo{author}{H.~M. Jaeger}, \bibinfo{author}{S.~R. Nagel},
  \bibinfo{author}{R.~P. Behringer}, \bibinfo{title}{Granular solids, liquids,
  and gases}, \bibinfo{journal}{Reviews of Modern Physics}
  \bibinfo{volume}{68}~(\bibinfo{number}{4}) (\bibinfo{year}{1996})
  \bibinfo{pages}{1259}.

\bibitem[{Yang(1998)}]{yang1998fluidization}
\bibinfo{author}{W.-C. Yang}, \bibinfo{title}{Fluidization, Solids Handling,
  and Processing: Industrial Applications}, \bibinfo{publisher}{Elsevier},
  \bibinfo{year}{1998}.

\bibitem[{Andreotti et~al.(2013)Andreotti, Forterre, and
  Pouliquen}]{andreotti2013granular}
\bibinfo{author}{B.~Andreotti}, \bibinfo{author}{Y.~Forterre},
  \bibinfo{author}{O.~Pouliquen}, \bibinfo{title}{Granular media: between fluid
  and solid}, \bibinfo{publisher}{Cambridge University Press},
  \bibinfo{year}{2013}.

\bibitem[{Goldhirsch(2003)}]{goldhirsch2003rapid}
\bibinfo{author}{I.~Goldhirsch}, \bibinfo{title}{Rapid granular flows},
  \bibinfo{journal}{Annual review of fluid mechanics}
  \bibinfo{volume}{35}~(\bibinfo{number}{1}) (\bibinfo{year}{2003})
  \bibinfo{pages}{267--293}.

\bibitem[{Aranson and Tsimring(2006)}]{aranson2006patterns}
\bibinfo{author}{I.~S. Aranson}, \bibinfo{author}{L.~S. Tsimring},
  \bibinfo{title}{Patterns and collective behavior in granular media:
  Theoretical concepts}, \bibinfo{journal}{Reviews of modern physics}
  \bibinfo{volume}{78}~(\bibinfo{number}{2}) (\bibinfo{year}{2006})
  \bibinfo{pages}{641}.

\bibitem[{Luding(2009)}]{luding2009towards}
\bibinfo{author}{S.~Luding}, \bibinfo{title}{Towards dense, realistic granular
  media in 2D}, \bibinfo{journal}{Nonlinearity}
  \bibinfo{volume}{22}~(\bibinfo{number}{12}) (\bibinfo{year}{2009})
  \bibinfo{pages}{R101--R146}.

\bibitem[{Kadanoff(1999)}]{kadanoff1999built}
\bibinfo{author}{L.~P. Kadanoff}, \bibinfo{title}{Built upon sand: Theoretical
  ideas inspired by granular flows}, \bibinfo{journal}{Reviews of Modern
  Physics} \bibinfo{volume}{71}~(\bibinfo{number}{1}) (\bibinfo{year}{1999})
  \bibinfo{pages}{435}.

\bibitem[{Brilliantov and P{\"o}schel(2004)}]{brilliantov2004kinetic}
\bibinfo{author}{N.~V. Brilliantov}, \bibinfo{author}{T.~P{\"o}schel},
  \bibinfo{title}{Kinetic theory of granular gases}, \bibinfo{publisher}{Oxford
  University Press}, \bibinfo{year}{2004}.

\bibitem[{P{\"o}schel and Schwager(2005)}]{poschel2005computational}
\bibinfo{author}{T.~P{\"o}schel}, \bibinfo{author}{T.~Schwager},
  \bibinfo{title}{Computational granular dynamics},
  \bibinfo{publisher}{Springer}, \bibinfo{year}{2005}.

\bibitem[{Goldhirsch(1999)}]{goldhirsch1999scales}
\bibinfo{author}{I.~Goldhirsch}, \bibinfo{title}{Scales and kinetics of
  granular flows}, \bibinfo{journal}{Chaos}
  \bibinfo{volume}{9}~(\bibinfo{number}{3}) (\bibinfo{year}{1999})
  \bibinfo{pages}{659--672}.

\bibitem[{Goldhirsch and Zanetti(1993)}]{goldhirsch1993clustering}
\bibinfo{author}{I.~Goldhirsch}, \bibinfo{author}{G.~Zanetti},
  \bibinfo{title}{Clustering instability in dissipative gases},
  \bibinfo{journal}{Physical review letters}
  \bibinfo{volume}{70}~(\bibinfo{number}{11}) (\bibinfo{year}{1993})
  \bibinfo{pages}{1619}.

\bibitem[{Campbell(1993)}]{campbell1993boundary}
\bibinfo{author}{C.~S. Campbell}, \bibinfo{title}{Boundary interactions for
  two-dimensional granular flows. Part 2. Roughened boundaries},
  \bibinfo{journal}{Journal of Fluid Mechanics} \bibinfo{volume}{247}
  (\bibinfo{year}{1993}) \bibinfo{pages}{137--156}.

\bibitem[{Galvin et~al.(2007)Galvin, Hrenya, and Wildman}]{galvin2007role}
\bibinfo{author}{J.~Galvin}, \bibinfo{author}{C.~Hrenya},
  \bibinfo{author}{R.~Wildman}, \bibinfo{title}{On the role of the Knudsen
  layer in rapid granular flows}, \bibinfo{journal}{Journal of Fluid Mechanics}
  \bibinfo{volume}{585} (\bibinfo{year}{2007}) \bibinfo{pages}{73--92}.

\bibitem[{Cundall and Strack(1979)}]{cundall1979discrete}
\bibinfo{author}{P.~A. Cundall}, \bibinfo{author}{O.~D. Strack},
  \bibinfo{title}{A discrete numerical model for granular assemblies},
  \bibinfo{journal}{Geotechnique} \bibinfo{volume}{29}~(\bibinfo{number}{1})
  (\bibinfo{year}{1979}) \bibinfo{pages}{47--65}.

\bibitem[{Wadsworth and Erwin(1990)}]{wadsworthone1990hybrid}
\bibinfo{author}{D.~Wadsworth}, \bibinfo{author}{Erwin},
  \bibinfo{title}{One-dimensional hybrid continuum/particle simulation approach
  for rarefied hypersonic flows}, \bibinfo{journal}{AIAA Paper}
  (\bibinfo{year}{1990}) \bibinfo{pages}{90--1690}.

\bibitem[{Kalweit and Drikakis(2008)}]{kalweit2008multiscale}
\bibinfo{author}{M.~Kalweit}, \bibinfo{author}{D.~Drikakis},
  \bibinfo{title}{Multiscale methods for micro/nano flows and materials},
  \bibinfo{journal}{Journal of Computational and Theoretical Nanoscience}
  \bibinfo{volume}{5}~(\bibinfo{number}{9}) (\bibinfo{year}{2008})
  \bibinfo{pages}{1923--1938}.

\bibitem[{Mohamed and Mohamad(2010)}]{mohamed2010review}
\bibinfo{author}{K.~Mohamed}, \bibinfo{author}{A.~Mohamad}, \bibinfo{title}{A
  review of the development of hybrid atomistic--continuum methods for dense
  fluids}, \bibinfo{journal}{Microfluidics and Nanofluidics}
  \bibinfo{volume}{8}~(\bibinfo{number}{3}) (\bibinfo{year}{2010})
  \bibinfo{pages}{283--302}.

\bibitem[{Miller and Tadmor(2009)}]{miller2009unified}
\bibinfo{author}{R.~E. Miller}, \bibinfo{author}{E.~Tadmor}, \bibinfo{title}{A
  unified framework and performance benchmark of fourteen multiscale
  atomistic/continuum coupling methods}, \bibinfo{journal}{Modelling and
  Simulation in Materials Science and Engineering}
  \bibinfo{volume}{17}~(\bibinfo{number}{5}) (\bibinfo{year}{2009})
  \bibinfo{pages}{053001}.

\bibitem[{Wijesinghe and Hadjiconstantinou(2004)}]{wijesinghe2004discussion}
\bibinfo{author}{H.~S. Wijesinghe}, \bibinfo{author}{N.~G. Hadjiconstantinou},
  \bibinfo{title}{Discussion of hybrid atomistic-continuum methods for
  multiscale hydrodynamics}, \bibinfo{journal}{International Journal for
  Multiscale Computational Engineering}
  \bibinfo{volume}{2}~(\bibinfo{number}{2}) (\bibinfo{year}{2004})
  \bibinfo{pages}{189}.

\bibitem[{Lun et~al.(1984)Lun, Savage, Jeffrey, and Chepurniy}]{lun1984kinitic}
\bibinfo{author}{C.~Lun}, \bibinfo{author}{S.~Savage},
  \bibinfo{author}{D.~Jeffrey}, \bibinfo{author}{N.~Chepurniy},
  \bibinfo{title}{Kinitic theories for granular flow: inelastic particles in
  Coutte flow and slighly inelastic particles in a general flowfield},
  \bibinfo{journal}{J. Fluid. Mech.}
  \bibinfo{volume}{140}~(\bibinfo{number}{0}) (\bibinfo{year}{1984})
  \bibinfo{pages}{223--256}.

\bibitem[{Kuipers et~al.(1993)Kuipers, Van~Duin, Van~Beckum, and
  Van~Swaaij}]{kuipers1993computer}
\bibinfo{author}{J.~A.~M. Kuipers}, \bibinfo{author}{K.~J. Van~Duin},
  \bibinfo{author}{F.~P.~H. Van~Beckum}, \bibinfo{author}{W.~P.~M. Van~Swaaij},
  \bibinfo{title}{Computer simulation of the hydrodynamics of a two-dimensional
  gas-fluidized bed}, \bibinfo{journal}{Computers \& chemical engineering}
  \bibinfo{volume}{17}~(\bibinfo{number}{8}) (\bibinfo{year}{1993})
  \bibinfo{pages}{839--858}.

\bibitem[{Van~der Hoef et~al.(2006)Van~der Hoef, Ye, van Sint~Annaland,
  Andrews, Sundaresan, and Kuipers}]{van2006multiscale}
\bibinfo{author}{M.~A. Van~der Hoef}, \bibinfo{author}{M.~Ye},
  \bibinfo{author}{M.~van Sint~Annaland}, \bibinfo{author}{A.~T. Andrews},
  \bibinfo{author}{S.~Sundaresan}, \bibinfo{author}{J.~A.~M. Kuipers},
  \bibinfo{title}{Multiscale modeling of gas-fluidized beds},
  \bibinfo{journal}{Advances in Chemical Engineering} \bibinfo{volume}{31}
  (\bibinfo{year}{2006}) \bibinfo{pages}{65--149}.

\bibitem[{Johnson and Jackson(1987)}]{johnson1987frictional}
\bibinfo{author}{P.~C. Johnson}, \bibinfo{author}{R.~Jackson},
  \bibinfo{title}{Frictional--collisional constitutive relations for granular
  materials, with application to plane shearing}, \bibinfo{journal}{Journal of
  Fluid Mechanics} \bibinfo{volume}{176} (\bibinfo{year}{1987})
  \bibinfo{pages}{67--93}.

\bibitem[{Ye et~al.(2004)Ye, Van~der Hoef, and Kuipers}]{ye2004numerical}
\bibinfo{author}{M.~Ye}, \bibinfo{author}{M.~Van~der Hoef},
  \bibinfo{author}{J.~Kuipers}, \bibinfo{title}{A numerical study of
  fluidization behavior of Geldart A particles using a discrete particle
  model}, \bibinfo{journal}{Powder technology}
  \bibinfo{volume}{139}~(\bibinfo{number}{2}) (\bibinfo{year}{2004})
  \bibinfo{pages}{129--139}.

\bibitem[{Garz{\'o} and Dufty(1999)}]{garzo1999dense}
\bibinfo{author}{V.~Garz{\'o}}, \bibinfo{author}{J.~Dufty},
  \bibinfo{title}{Dense fluid transport for inelastic hard spheres},
  \bibinfo{journal}{Physical Review E}
  \bibinfo{volume}{59}~(\bibinfo{number}{5}) (\bibinfo{year}{1999})
  \bibinfo{pages}{5895}.

\bibitem[{Li and Benyahia(2012)}]{li2012revisiting}
\bibinfo{author}{T.~Li}, \bibinfo{author}{S.~Benyahia},
  \bibinfo{title}{Revisiting Johnson and Jackson boundary conditions for
  granular flows}, \bibinfo{journal}{AIChE journal}
  \bibinfo{volume}{58}~(\bibinfo{number}{7}) (\bibinfo{year}{2012})
  \bibinfo{pages}{2058--2068}.

\bibitem[{Soleimani et~al.(2015)Soleimani, Schneiderbauer, and
  Pirker}]{soleimani2015comparison}
\bibinfo{author}{A.~Soleimani}, \bibinfo{author}{S.~Schneiderbauer},
  \bibinfo{author}{S.~Pirker}, \bibinfo{title}{A comparison for different
  wall-boundary conditions for kinetic theory based two-fluid models},
  \bibinfo{journal}{International Journal of Multiphase Flow}
  \bibinfo{volume}{71} (\bibinfo{year}{2015}) \bibinfo{pages}{94--97}.

\bibitem[{Di~Renzo and Di~Maio(2004)}]{di2004comparison}
\bibinfo{author}{A.~Di~Renzo}, \bibinfo{author}{F.~P. Di~Maio},
  \bibinfo{title}{Comparison of contact-force models for the simulation of
  collisions in DEM-based granular flow codes}, \bibinfo{journal}{Chemical
  Engineering Science} \bibinfo{volume}{59}~(\bibinfo{number}{3})
  (\bibinfo{year}{2004}) \bibinfo{pages}{525--541}.

\bibitem[{Schwager and P{\"o}schel(2007)}]{schwager2007coefficient}
\bibinfo{author}{T.~Schwager}, \bibinfo{author}{T.~P{\"o}schel},
  \bibinfo{title}{Coefficient of restitution and linear--dashpot model
  revisited}, \bibinfo{journal}{Granular Matter}
  \bibinfo{volume}{9}~(\bibinfo{number}{6}) (\bibinfo{year}{2007})
  \bibinfo{pages}{465--469}.

\bibitem[{O’Connell and Thompson(1995)}]{o1995molecular}
\bibinfo{author}{S.~T. O’Connell}, \bibinfo{author}{P.~A. Thompson},
  \bibinfo{title}{Molecular dynamics--continuum hybrid computations: a tool for
  studying complex fluid flows}, \bibinfo{journal}{Physical Review E}
  \bibinfo{volume}{52}~(\bibinfo{number}{6}) (\bibinfo{year}{1995})
  \bibinfo{pages}{R5792}.

\bibitem[{Cosden and Lukes(2013)}]{cosden2013hybrid}
\bibinfo{author}{I.~A. Cosden}, \bibinfo{author}{J.~R. Lukes},
  \bibinfo{title}{A hybrid atomistic--continuum model for fluid flow using
  LAMMPS and OpenFOAM}, \bibinfo{journal}{Computer Physics Communications}
  \bibinfo{volume}{184}~(\bibinfo{number}{8}) (\bibinfo{year}{2013})
  \bibinfo{pages}{1958--1965}.

\bibitem[{Nie et~al.(2004)Nie, Chen, Robbins et~al.}]{nie2004continuum}
\bibinfo{author}{X.~Nie}, \bibinfo{author}{S.~Chen},
  \bibinfo{author}{M.~Robbins}, et~al., \bibinfo{title}{A continuum and
  molecular dynamics hybrid method for micro-and nano-fluid flow},
  \bibinfo{journal}{Journal of Fluid Mechanics} \bibinfo{volume}{500}
  (\bibinfo{year}{2004}) \bibinfo{pages}{55--64}.

\bibitem[{Van~den Akker(2010)}]{van2010particle}
\bibinfo{author}{E.~Van~den Akker}, \bibinfo{title}{Particle-based Evaporation
  Models and Wall Interaction for Microchannel Cooling}, Ph.D. thesis,
  \bibinfo{school}{PhD Thesis Eindhoven University of Technology},
  \bibinfo{year}{2010}.

\bibitem[{Markesteijn(2011)}]{markesteijn2011connecting}
\bibinfo{author}{A.~P. Markesteijn}, \bibinfo{title}{Connecting molecular
  dynamics and computational fluid dynamics}, Ph.D. thesis,
  \bibinfo{school}{PhD Thesis Delft University of Technology},
  \bibinfo{year}{2011}.

\bibitem[{Chen and Wang(2015)}]{chen2015hybrid}
\bibinfo{author}{X.~Chen}, \bibinfo{author}{J.~Wang}, \bibinfo{title}{Hybrid
  Discrete-continuum Model for Granular Flow}, \bibinfo{journal}{Procedia
  Engineering} \bibinfo{volume}{102} (\bibinfo{year}{2015})
  \bibinfo{pages}{661--667}.

\bibitem[{Benenati and Brosilow(1962)}]{benenati1962void}
\bibinfo{author}{R.~Benenati}, \bibinfo{author}{C.~Brosilow},
  \bibinfo{title}{Void fraction distribution in beds of spheres},
  \bibinfo{journal}{AIChE Journal} \bibinfo{volume}{8}~(\bibinfo{number}{3})
  (\bibinfo{year}{1962}) \bibinfo{pages}{359--361}.

\bibitem[{Hartkamp et~al.(2012)Hartkamp, Ghosh, Weinhart, and
  Luding}]{hartkamp2012study}
\bibinfo{author}{R.~Hartkamp}, \bibinfo{author}{A.~Ghosh},
  \bibinfo{author}{T.~Weinhart}, \bibinfo{author}{S.~Luding}, \bibinfo{title}{A
  study of the anisotropy of stress in a fluid confined in a nanochannel},
  \bibinfo{journal}{The Journal of chemical physics}
  \bibinfo{volume}{137}~(\bibinfo{number}{4}) (\bibinfo{year}{2012})
  \bibinfo{pages}{044711}.

\bibitem[{Sederman et~al.(2001)Sederman, Alexander, and
  Gladden}]{sederman2001structure}
\bibinfo{author}{A.~Sederman}, \bibinfo{author}{P.~Alexander},
  \bibinfo{author}{L.~Gladden}, \bibinfo{title}{Structure of packed beds probed
  by magnetic resonance imaging}, \bibinfo{journal}{Powder Technology}
  \bibinfo{volume}{117}~(\bibinfo{number}{3}) (\bibinfo{year}{2001})
  \bibinfo{pages}{255--269}.

\bibitem[{Goldhirsch(2010)}]{goldhirsch2010stress}
\bibinfo{author}{I.~Goldhirsch}, \bibinfo{title}{Stress, stress asymmetry and
  couple stress: from discrete particles to continuous fields},
  \bibinfo{journal}{Granular Matter} \bibinfo{volume}{12}~(\bibinfo{number}{3})
  (\bibinfo{year}{2010}) \bibinfo{pages}{239--252}.

\bibitem[{Tsuji et~al.(2013)Tsuji, Narita, and Tanaka}]{tsuji2013effect}
\bibinfo{author}{T.~Tsuji}, \bibinfo{author}{E.~Narita},
  \bibinfo{author}{T.~Tanaka}, \bibinfo{title}{Effect of a wall on flow with
  dense particles}, \bibinfo{journal}{Advanced Powder Technology}
  \bibinfo{volume}{24}~(\bibinfo{number}{2}) (\bibinfo{year}{2013})
  \bibinfo{pages}{565--574}.

\bibitem[{Wang et~al.(1997)Wang, Jackson, and
  Sundaresan}]{wang1997instabilities}
\bibinfo{author}{C.-H. Wang}, \bibinfo{author}{R.~Jackson},
  \bibinfo{author}{S.~Sundaresan}, \bibinfo{title}{Instabilities of fully
  developed rapid flow of a granular material in a channel},
  \bibinfo{journal}{Journal of Fluid Mechanics} \bibinfo{volume}{342}
  (\bibinfo{year}{1997}) \bibinfo{pages}{179--197}.

\bibitem[{Liu and Glasser(2006)}]{liu2006parametric}
\bibinfo{author}{X.~Liu}, \bibinfo{author}{B.~J. Glasser}, \bibinfo{title}{A
  parametric investigation of gas-particle flow in a vertical duct},
  \bibinfo{journal}{AIChE journal} \bibinfo{volume}{52}~(\bibinfo{number}{3})
  (\bibinfo{year}{2006}) \bibinfo{pages}{940--956}.

\bibitem[{De~Fabritiis et~al.(2006)De~Fabritiis, Delgado-Buscalioni, and
  Coveney}]{de2006multiscale}
\bibinfo{author}{G.~De~Fabritiis}, \bibinfo{author}{R.~Delgado-Buscalioni},
  \bibinfo{author}{P.~Coveney}, \bibinfo{title}{Multiscale modeling of liquids
  with molecular specificity}, \bibinfo{journal}{Physical review letters}
  \bibinfo{volume}{97}~(\bibinfo{number}{13}) (\bibinfo{year}{2006})
  \bibinfo{pages}{134501}.

\bibitem[{Markesteijn et~al.(2014)Markesteijn, Karabasov, Scukins, Nerukh,
  Glotov, and Goloviznin}]{markesteijn2014concurrent}
\bibinfo{author}{A.~Markesteijn}, \bibinfo{author}{S.~Karabasov},
  \bibinfo{author}{A.~Scukins}, \bibinfo{author}{D.~Nerukh},
  \bibinfo{author}{V.~Glotov}, \bibinfo{author}{V.~Goloviznin},
  \bibinfo{title}{Concurrent multiscale modelling of atomistic and hydrodynamic
  processes in liquids}, \bibinfo{journal}{Philosophical Transactions of the
  Royal Society of London A: Mathematical, Physical and Engineering Sciences}
  \bibinfo{volume}{372}~(\bibinfo{number}{2021}) (\bibinfo{year}{2014})
  \bibinfo{pages}{20130379}.

\bibitem[{Vescovi et~al.(2014)Vescovi, Berzi, Richard, and
  Brodu}]{vescovi2014plane}
\bibinfo{author}{D.~Vescovi}, \bibinfo{author}{D.~Berzi},
  \bibinfo{author}{P.~Richard}, \bibinfo{author}{N.~Brodu},
  \bibinfo{title}{Plane shear flows of frictionless spheres: Kinetic theory and
  3D soft-sphere discrete element method simulations},
  \bibinfo{journal}{Physics of Fluids (1994-present)}
  \bibinfo{volume}{26}~(\bibinfo{number}{5}) (\bibinfo{year}{2014})
  \bibinfo{pages}{053305}.

\bibitem[{Almaz{\'a}n et~al.(2013)Almaz{\'a}n, Carrillo, Salue{\~n}a,
  Garz{\'o}, and P{\"o}schel}]{almazan2013numerical}
\bibinfo{author}{L.~Almaz{\'a}n}, \bibinfo{author}{J.~A. Carrillo},
  \bibinfo{author}{C.~Salue{\~n}a}, \bibinfo{author}{V.~Garz{\'o}},
  \bibinfo{author}{T.~P{\"o}schel}, \bibinfo{title}{A numerical study of the
  Navier--Stokes transport coefficients for two-dimensional granular
  hydrodynamics}, \bibinfo{journal}{New Journal of Physics}
  \bibinfo{volume}{15}~(\bibinfo{number}{4}) (\bibinfo{year}{2013})
  \bibinfo{pages}{043044}.

\bibitem[{Wang et~al.(2013)Wang, van~der Hoef, and
  Kuipers}]{wang2013comparison}
\bibinfo{author}{J.~Wang}, \bibinfo{author}{M.~van~der Hoef},
  \bibinfo{author}{J.~Kuipers}, \bibinfo{title}{Comparison of Two-Fluid and
  Discrete Particle Modeling of Dense Gas-Particle Flows in Gas-Fluidized
  Beds}, \bibinfo{journal}{Chemie Ingenieur Technik}
  \bibinfo{volume}{85}~(\bibinfo{number}{3}) (\bibinfo{year}{2013})
  \bibinfo{pages}{290--298}.

\bibitem[{Mitrano et~al.(2014)Mitrano, Zenk, Benyahia, Galvin, Dahl, and
  Hrenya}]{mitrano2014kinetic}
\bibinfo{author}{P.~P. Mitrano}, \bibinfo{author}{J.~R. Zenk},
  \bibinfo{author}{S.~Benyahia}, \bibinfo{author}{J.~E. Galvin},
  \bibinfo{author}{S.~R. Dahl}, \bibinfo{author}{C.~M. Hrenya},
  \bibinfo{title}{Kinetic-theory predictions of clustering instabilities in
  granular flows: beyond the small-Knudsen-number regime},
  \bibinfo{journal}{Journal of Fluid Mechanics} \bibinfo{volume}{738}
  (\bibinfo{year}{2014}) \bibinfo{pages}{R2}.

\bibitem[{Louge(2014)}]{louge2014surprising}
\bibinfo{author}{M.~Louge}, \bibinfo{title}{The surprising relevance of a
  continuum description to granular clusters}, \bibinfo{journal}{Journal of
  Fluid Mechanics} \bibinfo{volume}{742} (\bibinfo{year}{2014})
  \bibinfo{pages}{1--4}.

\bibitem[{Khain(2007)}]{khain2007hydrodynamics}
\bibinfo{author}{E.~Khain}, \bibinfo{title}{Hydrodynamics of fluid-solid
  coexistence in dense shear granular flow}, \bibinfo{journal}{Physical Review
  E} \bibinfo{volume}{75}~(\bibinfo{number}{5}) (\bibinfo{year}{2007})
  \bibinfo{pages}{051310}.

\end{thebibliography}
\end{document}